\documentclass{iaus}
\usepackage{graphicx}

\title[A Herschel study of PNe] %% give here short title %%
{A Herschel study of Planetary Nebulae}

\author[G. C. Van de Steene et al.]   %% give here short author list %%
{G. C. Van de Steene$^1$, K. M. Exter$^2$, P. A. M. van Hoof$^1$, T. L. Lim$^3$, M. J. Barlow$^4$, M. Matsuura$^4$, T. Ueta$^5$, \and the MESS consortium}

\affiliation{$^1$Royal Observatory of Belgium, Ringlaan 3, B-1180 Brussels, Belgium 
\\[\affilskip]
$^2$ IvS, Katholieke Universiteit Leuven, Celestijnenlaan 200 D, B-3001 Leuven, Belgium \\[\affilskip]
$^3$ Space Science and Technology Dept., Rutherford Appleton Lab., Oxfordshire OX11 0QX, UK \\[\affilskip]
$^4$ Dept. of Physics \& Astronomy, Univ. College London, Gower St, London WC1E 6BT, UK \\[\affilskip]
$^5$ Dept. of Physics and Astronomy, Univ. of Denver, Mail Stop 6900, Denver, CO 80208, USA \\[\affilskip]
}

\pubyear{2012}
\volume{283}  %% insert here IAU Symposium No.
\pagerange{xxx--xx}
\jname{Planetary Nebulae: an Eye to the Future}
\editors{A.C. Editor, B.D. Editor \& C.E. Editor, eds.}
\begin{document}

\maketitle

\begin{abstract}

  We present Herschel PACS and SPIRE images of the dust shells around
  the planetary nebulae NGC~650, NGC~6853, and NGC~6720, as well as
  images showing the dust temperature in their shells. The latter show
  a rich structure, which indicates that internal extinction in the UV
  is important despite the highly evolved status of the nebulae.

\end{abstract}

\firstsection % if your document starts with a section,
\section{Introduction}

As part of the Herschel Guaranteed Time Key Project MESS (Mass loss of
Evolved StarS) (PI Martin Groenewegen) we have imaged a sample of
planetary nebulae (PNe) with the PACS (Poglitsch et al. 2010) and SPIRE
(Griffin et al. 2010) instruments on board of the Herschel satellite
(Pillbratt et al. 2010). A detailed description of the program can be
found in Groenewegen et al. (2011) and an overview of the Herschel
observations for PNe in van Hoof et al. (2012).

%\cite[Anders \& Zinner (1993)]{AndersZinner93} and 
%\cite[Ott (1993)]{Ott93}.

\section{Data Reduction}

All targets in this paper have been imaged in scan map mode. With PACS
we have obtained images in the 70 and 160~$\mu$m bands, with SPIRE in the 250,
350, and 500~$\mu$m bands.  PACS data were reduced up to level~1
within the data procession package HIPE (Ott 2010). The PACS images
were made with the code Scanamorphos (Roussel 2011). The SPIRE images
were reduced with the SPIRE pipeline. The images were convolved using
the appropriate convolution kernels of Aniano et al. (2011) and
rebinned to the pixel size of the longest wavelength image with flux
conservation. These images were background subtracted before the
ratios were taken. In order to convert the flux ratio images to
temperature maps we determined the theoretical flux ratio at a given
grain temperature by folding the grain emissivity of astronomical
silicate or graphite (Martin \& Rouleau 1991) with the PACS and SPIRE
filter transmission curves in HIPE using the procedure outlined in the
SPIRE Observer's Manual.  We interpolated the flux ratio as a function
of temperature for each pixel in the the flux ratio image to obtain
the temperature map. The temperature maps based on the ratio
PACS~70~/~PACS~160~$\mu$m differs from PACS~160~/~SPIRE~250~$\mu$m
 by about 10~K.  The reason for this difference needs to be
investigated further. Here we present the temperature maps based on the 
PACS~70~/~PACS~160~$\mu$m ratio images.

\section{Results}

\begin{figure}
\begin{center}
\includegraphics[width=\textwidth]{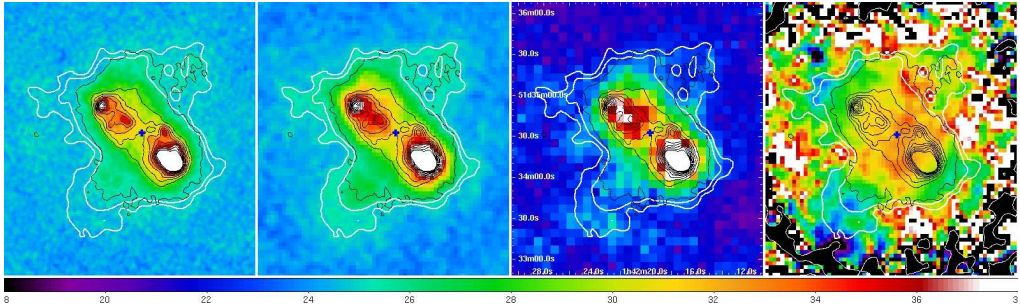} 
\caption{NGC~650, from left to right: PACS~70 and 160~$\mu$m,
  SPIRE~250~$\mu$m and the temperature map created from the
  PACS~70~/~160~$\mu$m ratio image. The black contours are of the
  PACS~70~$\mu$m inner region and the white contours of the fainter
  outer regions of the PACS~160~$\mu$m image. The blue cross marks the
  central star. The bar at the bottom shows the temperature scale.}
\label{fig1}
\end{center}
\end{figure}
\underline{\it NGC 650 (the Little Dumbbell)}:  
in the temperature map of Fig.\,\ref{fig1} the shadowing effect
of the torus is clearly visible: the regions at the outer edge of the
torus and beyond are clearly cooler than the dust in other
directions. The dust grains are primarily heated by UV photons, either
emitted by the central star, or diffuse emission from the gas (e.g.,
Ly$\alpha$ photons). Hence there must be substantial extinction of UV
photons inside the torus.
The measured flux values above 3~$\sigma$ are 6.54, 5.61 and 1.27~Jy
($\pm$~5\%), at 70, 160, and 250~$\mu$m respectively.

\underline{\it NGC 6853 (the Dumbbell)}:
in the temperature map of Fig.\,3 in van Hoof et al. (2012), we see a strong correlation between the high-density
regions and the colder dust. The hot patch towards the south appears
to be real and has no counterpart in the north. Presumably this is
material that is directly irradiated by the central star.
The measured flux values above 3~$\sigma$ are
83.45 and 56.0~Jy ($\pm$~5\%) at 70 and 160~$\mu$m respectively.

\underline{\it NGC 6720 (the Ring nebula)}:
in Fig.\,1 of van Hoof et al. (2012) we see that
the halo is clearly detected in all 3 bands. The temperature map shows
that the temperature is lower in the halo than inside the ring,
because it is shielded from the starlight by the dense ring.
The measured flux values above 3~$\sigma$ are 54.70, 29.18 and 13.70~Jy
($\pm$~5\%), at 70, 160, and 250~$\mu$m respectively.

The similarity between the optical H$\alpha$ and the FIR images,
indicate that the gas and cool dust are well mixed in these objects.
  
The detailed match between the H$_2$ emission and the FIR dust
emission, suggests the formation of H$_2$ on dust grains (van Hoof et
al. 2010).

\end{document}